\documentstyle[12pt, psfig]{article}
\def\bitem{\begin{itemize}}
\def\eitem{\end{itemize}}
\def\beq{\begin{equation}}
\def\eeq{\end{equation}}
\def\barr{\begin{eqnarray}}
\def\earr{\end{eqnarray}}
\def\bep{\epsilon}

\begin{document}

\begin{titlepage}
 
\begin{flushright}
IC/98/177 \\
hep-ph/9810381
\end{flushright}

\vspace{0.15truecm}
\begin{center}
\boldmath
\large\bf Information Content in $B \to VV$ Decays 
and the Angular Moments Method
\unboldmath
\end{center}
\vspace{0.3truecm}
\begin{center}
Amol Dighe  \footnote{Email : amol@ictp.trieste.it}\\[0.1cm]
{\sl The Abdus Salam International Centre for Theoretical Physics\\ 
34100 Trieste, Italy}\\[0.6cm]
\'Saunak Sen \footnote{Email : sen@stat.stanford.edu} \\[0.1cm]
{\sl Department of Statistics, 
        Stanford University, \\ Stanford, CA 94305, USA} \\[0.6cm]
\end{center}
\vspace{0.15cm}

\begin{abstract}
  
  The time-dependent angular distributions of decays of neutral $B$
  mesons into two vector mesons contain information about the
  lifetimes, mass differences, strong and weak phases, form factors,
  and CP violating quantities. A statistical analysis of the
  information content is performed by giving the ``information'' a
  quantitative meaning. It is shown that for some parameters of
  interest, the information content in time and angular measurements
  combined may be orders of magnitude more than the information from
  time measurements alone and hence the angular measurements are
  highly recommended. The method of angular moments is compared with
  the (maximum) likelihood method to find that it works almost as well
  in the region of interest for the one-angle distribution.  For the
  complete three-angle distribution, an estimate of possible
  statistical errors expected on the observables of interest is
  obtained.  It indicates that the three-angle distribution,
  unraveled by the method of angular moments, would be able to nail
  down many quantities of interest and will help in pointing 
  unambiguously to new physics.

\end{abstract}
 
\end{titlepage}

\setcounter{page}{1}

\section{Introduction}

Among the available methods for studying CP violation, the decay
modes of $B$ mesons into two vector mesons, both of which decay into
two particles each,  
are very promising mainly because of
the larger number of observables at one's disposal through
the angular distributions of the decays \cite{btovv}. 
A disadvantage of
having a large number of observables may be the difficulty in
separating them from one another because of the correlations between
them.  The method of angular moments \cite{dqstl, ddf1} helps in
extracting the observables from the angular distributions by using
judiciously chosen weighting functions. From the time evolutions of
these observables, it is then possible to extract the information
about the lifetimes, mass differences, strong and weak phases,
form factors, and CP violating quantities.

Here we will concentrate on the decays of the type 
$B \to V_1(\to X_1 Y_1) V_2 (\to X_2 Y_2)$, 
where $B$ is a neutral $B$ meson, $V_1$ and
$V_2$ are vector mesons and $X_1, X_2, Y_1, Y_2$ are the four final
state particles. We shall illustrate the technique by using the
particular decay $B_s \to J/\psi (\to \ell^+ \ell^-) \phi (\to K^+
K^-)$.  The other decay modes of the form $B \to VV$ might have
different angular distributions, and the method of angular moments
will need corresponding different weighting functions (which can
always be found \cite{ddf1}), but the observables in all these decay
modes are the same. In addition, $B_s \to J/\psi \phi$ decay holds the
promise of being able to measure the lifetimes of $B_s^H$ and $B_s^L$
separately, and, if this lifetime difference is sizeable (as estimated
in \cite{delta-gamma}), the prospect of measuring CP-violating
quantities even without tagging \cite{untagged}.  
By quantifying the information content in the data we can judge the
relative importance of the measurement of various possible
quantities.  The approach we have used to analyze the information
content may be used in modes of decay other than the one we have
considered here.   

Generally speaking, in any experiment the amount of information
obtained depends on
\begin{list}{$\bullet$}{\setlength{\itemsep}{-\parsep} 
\setlength{\topsep}{2pt}}
\item what quantities are recorded
\item what numerical summaries of the recorded data were used
\item the number of data points 
\item the parameter values governing the outcome of the experiment
\end{list}

Since only the first two are under the direct control of the
experimentalist, we will address those two issues in this paper.  We
argue that the expected information {\em per observation} available in
the $B_s \to J/\psi \phi$ decay
is substantially more when both time and angular information are used
instead of using the time information alone.  Moreover, we show that
the method of angular moments, when used to summarize and estimate the
parameters, is computationally easy to implement and efficient (in the
statistical sense) in extracting information from the data.

In Sec.~\ref{distr}, we give the angular distribution and the time
evolutions of the observables for the decay $B_s \to J/\psi \phi$.
The definition of information in the data about a parameter value 
that we will use is
standard in the statistical literature and will be described briefly in
Appendix~\ref{exp-info}. Sec.~\ref{sec:angular} outlines why the angular
information may be useful and then follows up with an analytic
investigation of the additional information in the transversity angle
over and above the time information.  In Sec.~\ref{eff}, we discuss
the efficiency of the method of angular moments by comparing it with
the the maximum likelihood method in the case of the transversity
angle distribution.  In Sec.~\ref{three-ang} we carry out a simulation
study of the method of angular moments for extracting the relevant
parameters from the three angle distribution.  
Sec.~\ref{concl} concludes.

\section{Angular Distributions and Time Evolutions of Observables}
\label{distr}

        The most general decay amplitude for $B \to VV$ takes
the form \cite{rosner,ddlr}
\beq\label{ampl}
A(B_q(t) \to V_1V_2) =
\frac{A_0(t)}{x}  {\bep}^{*L}_{V_1}
{\bep}^{*L}_{V_2} -
A_{\|}(t) {\bep}^{*T}_{V_1} \cdot
{\bep}^{*T}_{V_2} / \sqrt{2}  -
i A_{\perp}(t) {\bep}^*_{V_1} \times
{\bep}^*_{V_2} \cdot \hat{\bf p}_{V_2} /
                    \sqrt{2}~,
\eeq
where $x\equiv p_{V_1}\cdot p_{V_2}/(m_{V_1} m_{V_2})$ and
$\widehat{\bf p}_{V_2}$ is the unit vector along the direction of
motion of $V_2$ in the rest frame of $V_1$.
Here the time dependences originate from $B_q-\overline{B}_q$
mixing. In our notation only a $B_q$ meson is present at $t=0$.

For angles, we will use the same conventions as
in Ref.~\cite{ddlr}, i.e.\ $\phi$ moves in the $x$ direction in
the $J/\psi$ rest frame, the $z$  axis is perpendicular to
the decay plane of $\phi \to K^+ K^-$, and $p_y(K^+)
\geq 0$. The coordinates $(\theta, \varphi)$ describe the decay
direction of $l^+$ in the $J/ \psi$ rest frame and $\psi$ is
the angle made by $\vec p(K^+)$ with the $x$ axis in the $\phi$
rest frame. With this convention,
\barr
{\bf x} = {\bf p}_{\phi} , &
{\bf y} = \frac{ {\bf p}_{K^+} - {\bf p}_{\phi} ( {\bf p}_{\phi}
                                \cdot {\bf p}_{K^+} ) }
        { | {\bf p}_{K^+} - {\bf p}_{\phi} ( {\bf p}_{\phi}
                                \cdot {\bf p}_{K^+} ) | } , &
{\bf z} = {\bf x} \times {\bf y} , \nonumber \\
\sin \theta \;\; \cos \varphi =   {\bf p}_{\ell^+} \cdot  {\bf x}, &
\sin \theta \;\; \sin \varphi = {\bf p}_{\ell^+} \cdot  {\bf y}, &
\cos \theta =  {\bf p}_{\ell^+} \cdot {\bf z}~~.
\label{angdef}
\earr
Here boldface characters represent {\it  unit} 3-vectors and
everything is measured in the rest frame of $J/\psi$. Also
\beq
\cos \psi = - {\bf p}'_{K^+}
                        \cdot {\bf p}'_{J/\psi},
\label{psidef}
\eeq
where the primed quantities are {\it  unit vectors}
measured in the rest frame of $\phi$.

With this convention,
the three angle distribution is given by \cite{ddf1, ddlr}
$$
\frac{d^3 \Gamma [B_s(t) \to J/\psi (\to l^+ l^-) \phi (\to K^+
K^-)]}
{d \cos \theta~d \varphi~d \cos \psi}
\propto \frac{9}{32 \pi} \Bigl[~2 |A_0(t)|^2 \cos^2 \psi (1 - \sin^2
\theta
\cos^2 \varphi)
\hfill{ }
$$
$$
+ \sin^2 \psi \{ |A_\parallel(t)|^2  (1 - \sin^2 \theta \sin^2
\varphi)
+ |A_\perp(t)|^2 \sin^2 \theta - \mbox{ Im }(A_\parallel^*(t)
A_\perp(t))
\sin 2 \theta \sin \varphi \}~~~
$$
\beq \label{triple}
+\frac{1}{\sqrt{2}}\sin2\psi\{{\mbox{ Re }}(A_0^*(t) A_\parallel(t))
\sin^2
\theta
\sin2\varphi+\mbox{ Im }(A_0^*(t)A_\perp(t))\sin2\theta\cos\varphi~\}
~\Bigr]~~.
\eeq

The time evolutions of the coefficients of the angular terms are given
in Table~\ref{tab1}. Here $\Gamma_L$ and $\Gamma_H$ are the widths of
the light and heavy $B_s$ mass eigenstates, $B_s^L$ and $B_s^H$
respectively, and $\Delta m$ is the mass difference between them.
$\overline{\Gamma}$ is the average of $\Gamma_L$ and $\Gamma_H$.  Here
$\delta_1 \equiv Arg(A_{\|}^{\ast}(0) A_{\perp}(0)) $ and $\delta_2
\equiv Arg(A_0^{\ast}(0) A_{\perp}(0)) $ are the strong phases, and
$\delta \phi \approx 2 \lambda^2 \eta$ is related to an angle of a
(squashed) unitarity triangle \cite{dphi}, which is very small in the
standard model ($\approx 0.03$).
We will denote the values of $A_X(0)$ (where $X \in \{0, \|, \perp \}$)
simply as $A_X$ in the rest of the paper.

The `transversity angle' $\theta$ separates the CP-even and 
CP-odd decays. If we integrate over the remaining two angles
and include the time dependence explicitly,
the angular distribution in Eq.~(\ref{triple}) becomes

\beq
\label{time+angle}
\frac{d^2\Gamma}{d\cos\theta~ dt} \propto
(|A_0|^2 + |A_{\parallel}|^2)(1+ \cos^2 \theta) e^{-\Gamma_L t}
+ |A_{\perp}|^2 \sin^2 \theta ~e^{-\Gamma_H t}~~~,
\eeq
or, in the form of a normalized probability distribution,
\beq
\label{time-angle}
p(u,t|\beta, \Gamma_H, \Gamma_L) = 
        \frac{3}{8} \beta \Gamma_L (1+u^2) e^{-\Gamma_L t} +
        \frac{3}{4}  (1 - \beta) \Gamma_H (1-u^2) e^{-\Gamma_H t}~~, 
\eeq
where $u \equiv \cos \theta$ and 
\beq
\label{beta-def}
\beta \equiv \left( 1 + \frac{1}{2} \frac{\Gamma_L}{\Gamma_H}
        \frac{ |A_{\perp}|^2 }{|A_0|^2 + |A_{\parallel}|^2} 
        \right) ^{-1}~~. 
\eeq

The corresponding value of $\beta$ in the
$B_d \to J/\psi K^{\ast}$ mode is measured 
\cite{cdf-1,cleo} to be $0.93 \pm 0.03$, so 
with conservative estimates for the breaking of 
flavour SU(3) symmetry,
the value of $\beta$ is expected to lie between 0.8 and 1.0.

\section{Information in the transversity angle distribution}
\label{sec:angular}

By considering the case when the time ($t$) and transversity angle
($\theta$) measurements are available, we will argue, in this section,
that the additional information in $\theta$ is substantial and worth
the extra effort put into the angular measurements.  In Section 3.1 we
will explain what makes the estimation of $\Gamma_L - \Gamma_H$ hard
and why gathering the transversity angle data in addition to time is
attractive.  In Section 3.2 we will analyze the information content
analytically and determine the numerical magnitude of the information
gain.

\subsection{Why collect angular information?}
\label{subsec:why}

If the objective is to estimate $\Gamma_L - \Gamma_H$, which is the
difference between the reciprocal of mean lifetimes, one may ask: how
is it possible that the angular data is useful?

Let us first consider the estimation of the parameters when
we have only time information available.  In that case the
distribution of the lifetime ($t$) is given by
\beq
\label{time-only}
p(t|\beta, \Gamma_H, \Gamma_L) = \beta \Gamma_L e^{-\Gamma_L t} + (1 -
\beta) \Gamma_H e^{-\Gamma_H t}~~,  \eeq 
which is a {\em mixture} of the two lifetimes.  With probability
$\beta$ we observe the lifetime of a particle with mean lifetime
$1/\Gamma_L$ and with probability $(1-\beta)$ we observe the lifetime of
a particle with lifetime $1/\Gamma_H$.  The expectation of the observed
(mixed) lifetime is $\beta /\Gamma_L + (1{-}\beta)/\Gamma_H$.  Since
\begin{equation}
  \left( \frac{\beta}{\Gamma_L} + \frac{(1-\beta)}{\Gamma_H} \right)
  \times \Big( \beta \Gamma_L + (1{-}\beta) \Gamma_H \Big) = 1 - \beta
  (1{-}\beta) \left( \sqrt{\frac{\Gamma_H}{\Gamma_L}} - \sqrt{\frac{\Gamma_L}
  {\Gamma_H}} \right)^2 ~~,
\end{equation}
the derived parameter $\beta \Gamma_L + (1{-}\beta) \Gamma_H$ (which
we will later call $\theta_1$) is, to the first order (when the
$\Gamma$'s are close to each other), the reciprocal
of the expected mean observed lifetime.  Thus the estimation of $\beta
\Gamma_L + (1{-}\beta) \Gamma_H$ which is a ``mean'' parameter is not
hard even if we cannot ``guess'' the decay type.

If we knew what type of decay each time measurement was coming from,
then we could estimate $\Gamma_L$ and $\Gamma_H$ separately by using
the reciprocal of the sample mean lifetimes of the two kinds of
decays. We can then construct an estimate for $\Gamma_L -
\Gamma_H$.  Given enough observations of both types, we can get good
estimates for the difference of the lifetimes.  However, the identity
of the decay type is not known in real data and statistical procedures
have to, at least indirectly, guess it as well as possible from the
available data.  When the two component lifetimes are widely
different, then the observed lifetime is a good clue as to the
identity of the decay type.  However, when the lifetimes are close to
each other, the clues in the time signature alone are not decisive.

If only the transversity angle, $\theta$, is measured, then the
density of the data $u=\cos(\theta)$ is given by
\beq
\label{eq:only-angle}
 p(u|\beta) = \frac{3}{8} \beta (1 + u^2) +
\frac{3}{4} (1-\beta) (1 - u^2) ~~.
\eeq
The distribution of the angles is very dependent on the type of the
decay; therefore, by observing the angle alone one can have a fair
idea as to what kind of decay has been observed. 
This is why the angular
information is useful, even though it is not directly about lifetimes.

\begin{figure} [t]
\centerline{\psfig{figure=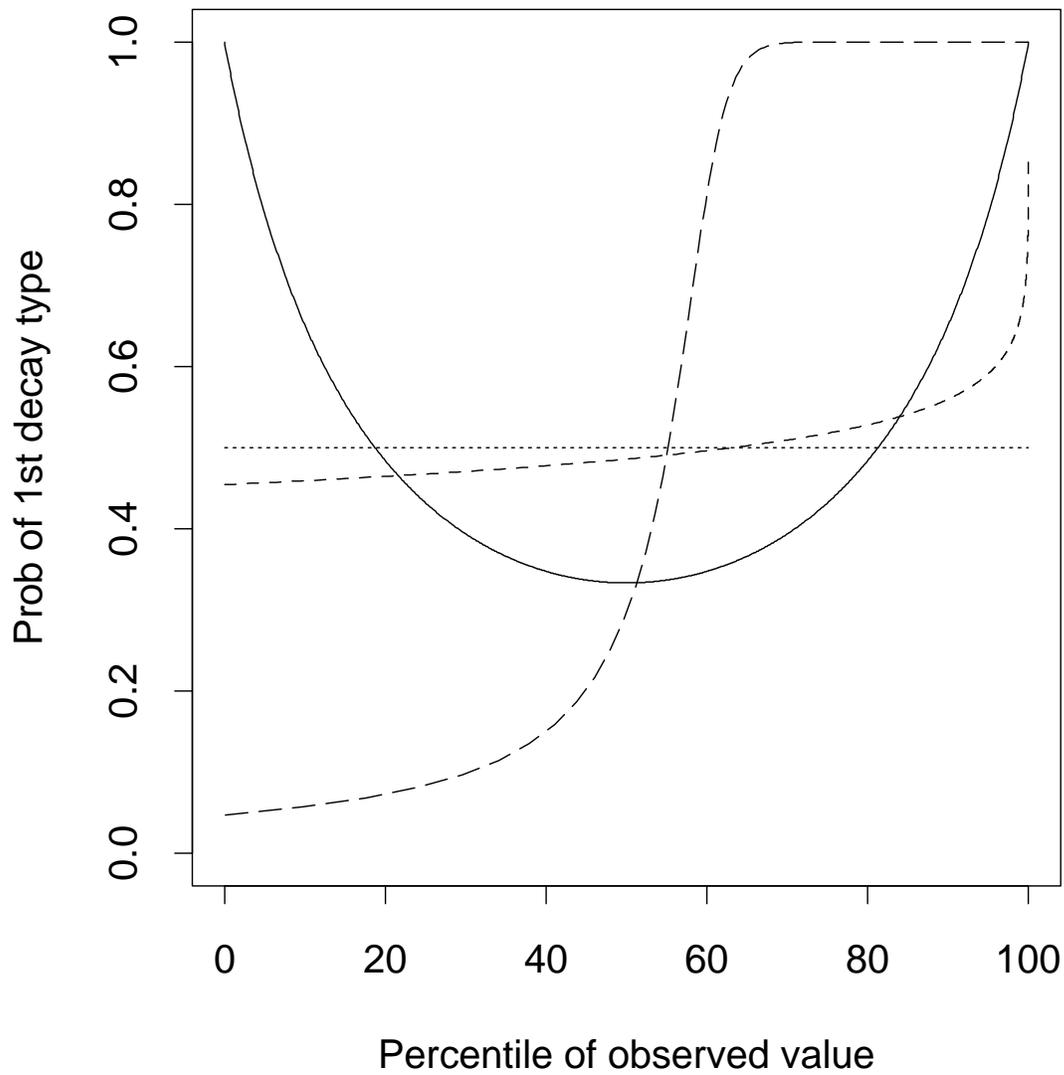,height=15cm}}
\caption{The ability to guess the decay type given 
        time or angular data.
\label{fig:guess} }
\end{figure}

As an illustrative example, consider the case when the two lifetimes 
are equally likely, i.e.
$\beta=0.5$ (see Fig.~\ref{fig:guess}). If $\Gamma_H/\Gamma_L=1$ and
only time measurements are available, then we have no way of
``guessing'' what kind of decay we have observed.  
This is reflected in the
fact that the {\em a priori} probability (the probability before
making the measurement) that the decay is of the first type is the
same as the {\em a posteriori} probability (the probability after
making the measurement) and equal to 0.5.  On the other hand, if the
decay widths are dramatically different, say $\Gamma_H/\Gamma_L=20$,
then the time measurement provides a good clue: if the time
observation is very large, then it is more likely that the decaying
particle is the one
with the smaller decay width and if the time measurement is very
small, then it is more likely that the decaying particle has 
the larger decay width.  

If only angular measurements are available, then when
$u=\cos(\theta)$ is very small or very large, there is a higher
chance that the particle measured was of the first type because it has
the angular distribution of $\frac{3}{8} (1 + u^2) $ which implies more
probability for large values of $|u|$.  Note that the power to
discriminate between the two kinds of decay by observing the
transversity angle is not affected by the ratio of the decay widths.
When we have {\em both} time and angular information, we will be able
to benefit from the information contained in both which will be at
least as much as the information in the angle alone.

The heuristic ideas above are graphically presented in
Fig.~\ref{fig:guess}.  The x-axis was chosen to be the the percentile
of the observed data (time or angle, as the case may be) so that we
can plot the different scenarios on the same scale.  Additionally the
plot has the desirable property that all points along the x-axis occur with
equal probability for all the four scenarios considered. (This is
because the percentile of the observed data is just 100 times the {\em
  probability integral transform}\footnote{ \label{foot:pit}If $X$ is
  a continuous random variable with density function $f(x)$, then the
  function $F(x)=\int_{-\infty}^{x} f(t) dt = P(X\leq x)$ defines the
  probability integral transform.  Then the random variable $F(X)$ is
  uniformly distributed over the interval $[0,1]$.} of the data point.)
Thus we can visually look at the four curves and compare how much they
deviate {\em away in either direction } from the line $y=0.5$ to get
an idea as to how well the data predicts the kind of decay.

The curve corresponding to $\Gamma_H/\Gamma_L=1.2$ when only time is
measured, is {\it closer} to the line $y=0.5$ than the curve
corresponding to when only angular measurements are taken.  This
implies that when the decay widths are close (for example when the
ratio is 1.2), the information in the angular data alone is
greater than that in the time data alone.  Only in extreme cases, such
as when the ratio of the decay widths is 20, can we predict well on
the basis of time alone.

\subsection{A theoretical investigation of information content}

In this subsection, we will consider the problem of extracting
information from a mixture of two distributions with the density of
observations of the form
\begin{equation}
p(x|\lambda_1, \lambda_2, \beta) = \beta g_1(x|\lambda_1) + (1-\beta)
g_2(x|\lambda_2).
\end{equation}
The data with or without angular information has this form ($x$
denotes the data from a single observation and may be a vector).
According to the equation above, the data comes from the distribution
$g_1(x|\lambda_1)$ with probability $\beta$ and from the distribution
$g_2(x|\lambda_2)$ with probability $(1-\beta)$.  This setup is more
general than the one we have, but it enables us to analyze the
phenomenon with greater clarity and it is also applicable to
data-collection scenarios other than $B \rightarrow VV$.  
For the $B \to VV$ case, 
$\lambda_1 = \Gamma_L$ and $\lambda_2 = \Gamma_H$.

When only time information is available,
\begin{equation}
  x \equiv t,\;\; g_1(x|\lambda_1) \equiv \lambda_1 \exp(-\lambda_1
 t),\;\;  g_2(x|\lambda_2) \equiv \lambda_2 \exp(-\lambda_2 t).
\end{equation}
When both time and angular information are available,
\begin{eqnarray}
  x & \equiv & (t,u), \nonumber\\ 
\label{eq:time-angle-1}
g_1(x|\lambda_1) & \equiv & \frac{3}{8} (1+u^2) \lambda_1
  \exp(-\lambda_1 t ), \\ 
\label{eq:time-angle-2}
g_2(x|\lambda_1) & \equiv & \frac{3}{4} (1-u^2) \lambda_2
  \exp(-\lambda_2 t ).
\end{eqnarray}
  When only time
information is collected, the functions $g_1(\cdot)$ and $g_2(\cdot)$
are the same.  When both time and angular information are recorded,
they will be different.

The expected information matrix (See Appendix A) for the parameters 
$\lambda = (\lambda_1,\lambda_2,\beta)$ can be found to be
\begin{eqnarray}
I(\lambda) & = & \left(
\begin{array}{ccc}
\beta^2 {\int} A^2 d\mu & \beta (1{-}\beta) {\int} AB d\mu & \beta {\int} AC
d\mu \\
\beta (1{-}\beta) {\int} AB d\mu & (1{-}\beta)^2 {\int} B^2 d\mu & (1{-}\beta)
{\int} BC d\mu \\
\beta {\int} AC d\mu & (1{-}\beta) {\int} BC d\mu & {\int} C^2 d\mu \\
\end{array}
\right)\\
 & = & \int [v, v]\, d\mu
\end{eqnarray}
where
$A = g_1^\prime(x|\lambda_1)$, $B = g_2^\prime(x|\lambda_2)$, $C=
g_1(x|\lambda_1) - g_2(x|\lambda_2)$, 
$$ v = \left(
  \begin{array}{c}
    \beta A \\
    (1-\beta) B \\
    C
  \end{array}
\right),$$
$[\cdot,\cdot]$ denotes outer product and ${\int} d\mu$ denotes
integration with respect to the measure $dx/p(x|\lambda_1, \lambda_2,
\beta)$.

If we are interested in the difference of the two $\lambda$'s then we
may be interested in the following derived parameterization of the
problem :
$$ \theta = \left(
\begin{array}{c}
\theta_1 \\
\theta_2 \\
\theta_3 \\
\end{array}
\right) = \left(
  \begin{array}{c}
    \beta \lambda_1 + (1{-}\beta) \lambda_2 \\
    \lambda_1 - \lambda_2 \\
    \beta \\
  \end{array} 
\right)~~. 
$$
In that case the information matrix for $\theta$ will be
\begin{eqnarray}
\lefteqn{I(\theta) = \int [w,w] \; d\mu }\\
& = &\left( 
{\scriptsize
\begin{array}{ccc}
{\int} (\beta A + (1{-}\beta) B)^2 d\mu & \ast &  \ast \\
\ast & (\beta (1 {-} \beta))^2 {\int} (A {-} B)^2 d\mu & \ast \\
\ast & \ast & {\int} ( (\lambda_2 - \lambda_1) (\beta  A + (1{-}\beta) B ) + C
)^2 d\mu \\
\end{array}
}
\right), \nonumber \\
& & 
\label{thetadef}
\end{eqnarray}
where
$$ w = \left(
  \begin{array}{c}
    \beta A + (1-\beta) B\\
    \beta(1-\beta)(A-B) \\
    (\lambda_2-\lambda_1)( \beta A +(1-\beta)B) + C
  \end{array}
\right).$$

The entries marked with a $\ast$ are important, they are omitted for
the sake of brevity since the qualitative features of the information
matrix are clear without them.  Careful inspection of the entries in
the above matrix in (\ref{thetadef}) reveals
some qualitative features of the dependence of the information content
in the data on the parameter values.  

The information on $\theta_2 = \lambda_1-\lambda_2$ is low when $A
\simeq B$ or if $\beta(1{-}\beta) \simeq 0$.  If $g_1 \equiv g_2$ and
$\lambda_1 \simeq \lambda_2$, then $A \simeq B$.  This is what happens
when we have only time data and the two $\Gamma$'s are close to each
other.  If $\lambda_1 \simeq \lambda_2$, but $g_1 \not\equiv g_2$,
then this problem does not occur.  In fact, if $g_1$ and $g_2$ are
very different functions then even if $\lambda_1 \simeq \lambda_2$, we
can recover information about $\lambda_1-\lambda_2$ from the data.
When both time {\em and} angle data are collected, the component
densities as in (\ref{eq:time-angle-1}) and (\ref{eq:time-angle-2})
are well-separated and there is a lot more information on $\Gamma_H -
\Gamma_L$ than what would have been with time data alone.  If most of
the observations are from one component of the mixture, the
information on $\lambda_1-\lambda_2$ is small, since $\beta (1 -
\beta) \simeq 0$.

The estimation of $\beta \Gamma_L + (1 {-} \beta) \Gamma_H$, on the
other hand, is not affected much by the separation of densities since
it is a ``mean'' parameter, as shown in Sec.~\ref{subsec:why}.

As mentioned in Appendix~\ref{exp-info}, the inverse of the expected
information matrix also gives the approximate variance matrix of the
maximum likelihood estimates in large samples.  If $I(\theta)$
is the expected information matrix from a single sample, $n I(\theta)$
is the expected information matrix based on a sample of size $n$.
Hence the diagonal elements of the matrix \beq \frac{V(\theta)}{n} =
\frac{1}{n} [I(\theta)]^{-1} \eeq will give us the approximate
variance, $V(\hat{\theta_i})$, of the maximum likelihood estimates,
$\hat{\theta_i}$ ($i=1,2,3$), in samples of size $n$ when $n$ is
large.

\begin{figure}[t]
\centerline{\psfig{figure=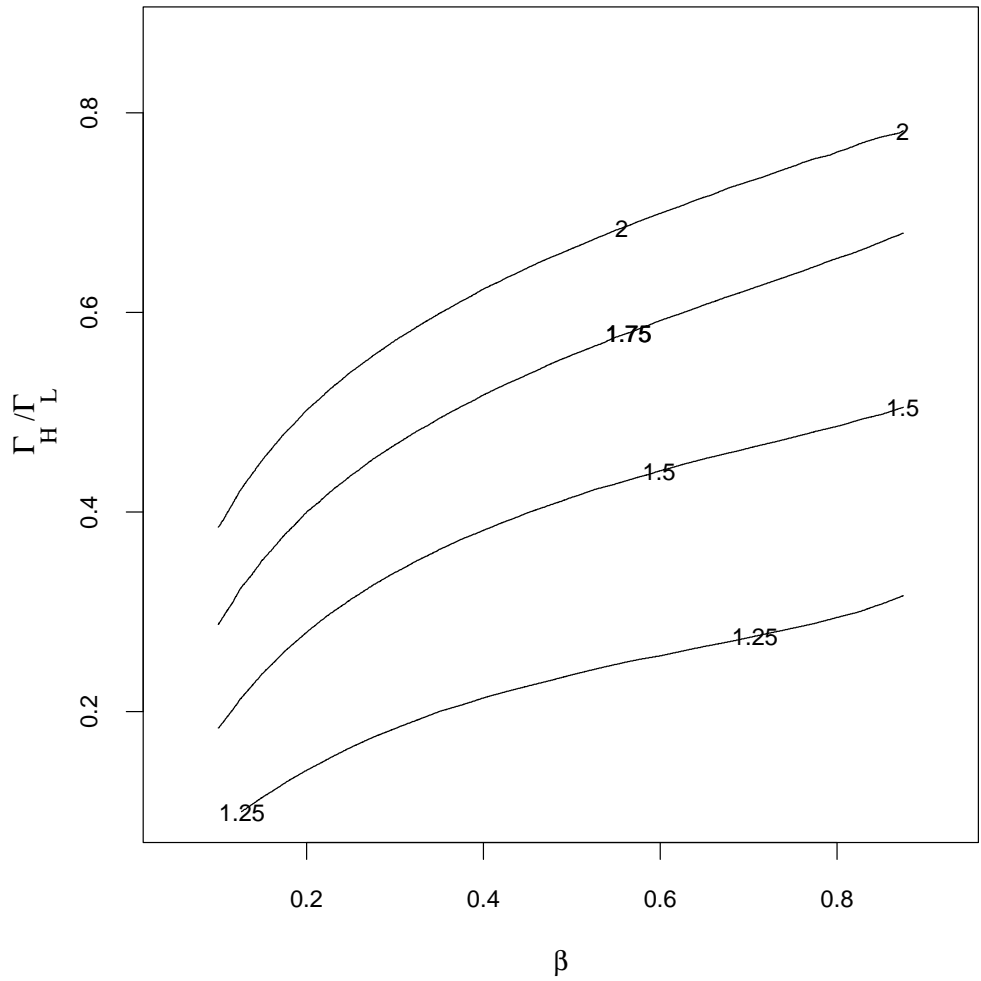,height=15cm}}
\caption{The ratio of the variances,
$V(\hat{\theta}_i(t)) / V(\hat{\theta}_i(u,t))$ of the estimates of
$\theta_1 = \beta \Gamma_L + (1-\beta) \Gamma_H$.
\label{fig:sumvar}}
\end{figure}

Let $\hat{\theta}_i(t)$ denote the maximum likelihood estimate of
$\theta_i$ given time data alone and let $\hat{\theta}_i(t,u)$ denote
the estimate using both time and angular information.  By calculating
the inverses of the corresponding expected information matrices, one
can calculate the ratios 
$$ \frac{V(\hat{\theta}_i(t))}{V(\hat{\theta}_i(t,u))},$$
for $i=1,2,3$.  Figs.~\ref{fig:sumvar}{-}\ref{fig:betavar} 
show the plots of
these ratios for
various values of $\beta$ and $\Gamma_H/\Gamma_L$.

\begin{figure}[t]
\centerline{\psfig{figure=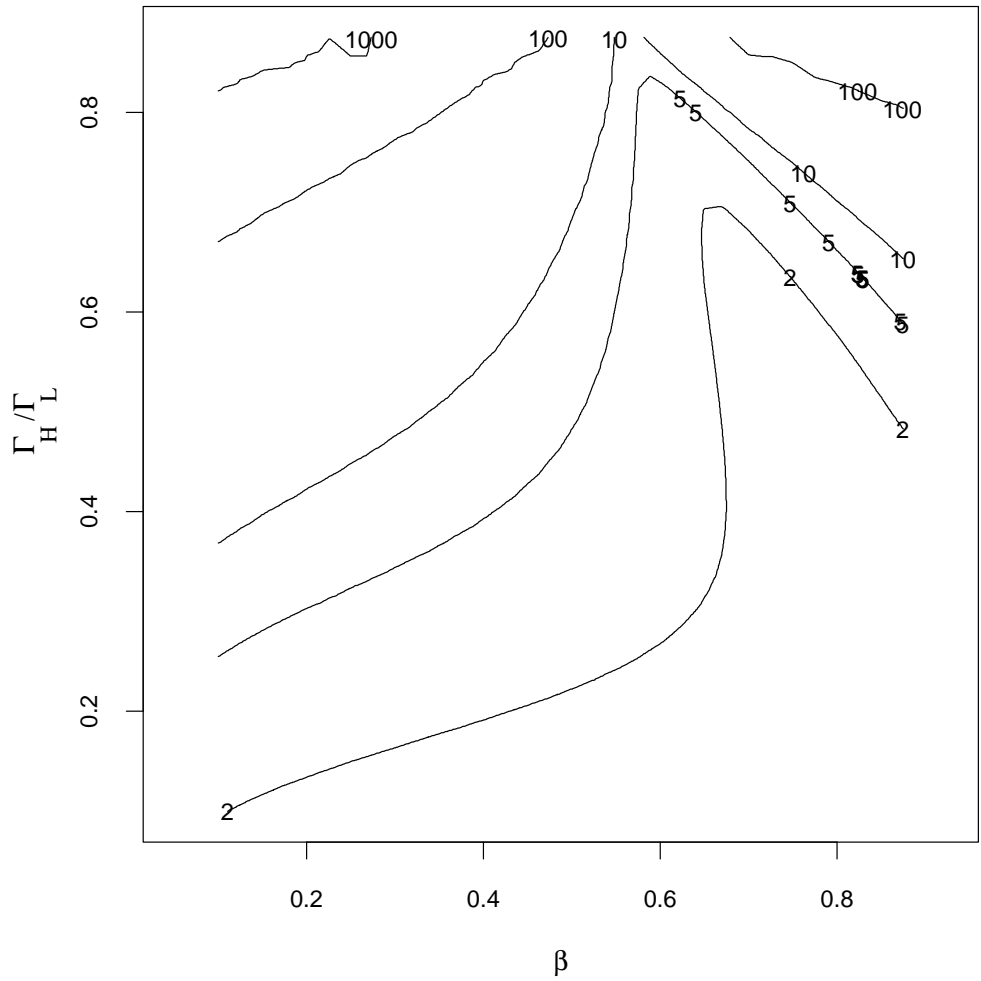,height=15cm}}
\caption{The ratio of the variances,
$V(\hat{\theta}_i(t)) / V(\hat{\theta}_i(u,t))$ of the estimates of
$\theta_2 = \Gamma_L - \Gamma_H$.
\label{fig:diffvar}}
\end{figure}

\begin{figure}[t]
\centerline{\psfig{figure=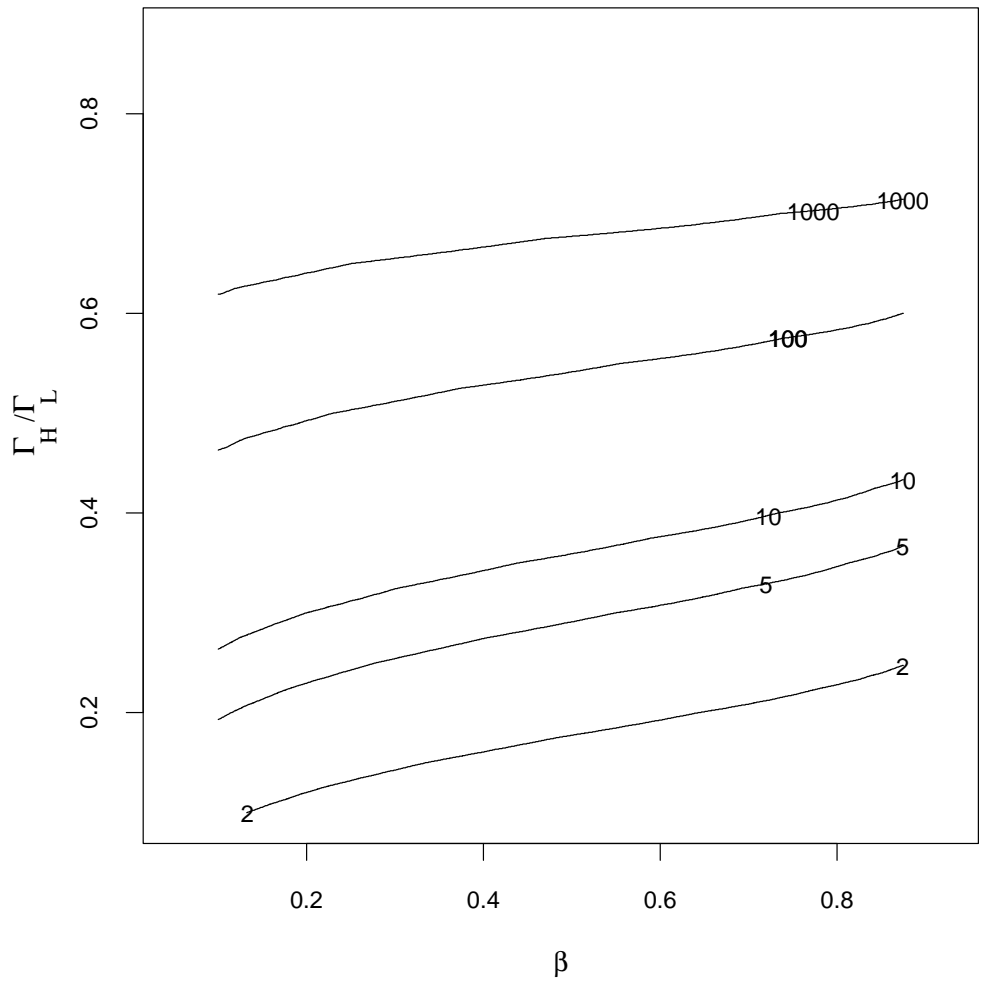,height=15cm}}
\caption{The ratio of the variances,
$V(\hat{\theta}_i(t)) / V(\hat{\theta}_i(u,t))$ of the estimates of
$\theta_3 = \beta$.
\label{fig:betavar}}
\end{figure}

Whereas most of the information about $\theta_1$ is indeed in the time
measurements as expected, it can be seen that the variance of the
parameters $\lambda_1 - \lambda_2$ and $\beta$ is orders of magnitude
higher if we have only the time information than if we had both the
time and angle information.  The physical region of interest lies
around $0.8 < \beta$ and $0.8 < \Gamma_H/ \Gamma_L$, where the
disparity in the two values of variances is striking.  The width of
confidence intervals is proportional to the standard error which is
the square root of the variance of the estimator.  Since the variance
of maximum likelihood estimates and the moment estimates are inversely
proportional to the number of data points, the ratio of the number of
data points needed to have confidence intervals of a given length,
with time data alone instead of time and angle data, is equal to the
levels of the contours in Figs.~\ref{fig:sumvar}{-}\ref{fig:betavar}.
Looking at the upper right hand corner of the plot in
Fig.~\ref{fig:diffvar}, we can see that with the inclusion of the
angular information, the sample sizes required for estimation of
$\Gamma_L-\Gamma_H$ to a desired level of accuracy will be smaller by a
factor of at least 10 times or even more than 100 times compared to
those required with the time information alone.

The analysis of the data using the angular information available is,
therefore, highly recommended.

\section{The method of angular moments for extracting information}
\label{eff}

Because of the optimality properties that the maximum likelihood
estimates enjoy in large samples, the method of maximum likelihood is
widely used.  The likelihood function, indeed, contains all the
information available in the data, and is the most efficient method
for summarizing the information in the data when the form of the
probabilistic model for the data is known \cite{coxandh}.  However,
there are some practical limitations to the likelihood method.  When
the number of parameters is large, exploring the likelihood surface is
problematic.  Finding the maximum is also difficult.  In addition, if
proper care is not taken, misleading results may be obtained (see for
example, \cite{thisted}, chapter on ``Non-Linear Statistical
Methods''), and when there are random errors in the measurement
process, the likelihood function may not be computable (see section
4.2).

We, therefore, propose the method of angular moments, which sacrifices
on some information (as compared to the likelihood function), but can
give consistent and reliable estimates of the parameters in a clear
way.

In the following section, we will show that in the case of one
angle distribution at least, the angular moments 
method is almost as efficient
as the maximum likelihood method. In
sections 4.2 and 4.3 we will discuss the effects of imperfections in
the measurement process on both the method of angular moments and the
likelihood method.

\subsection{The efficiency of the method of angular moments}

The method of angular moments is described in \cite{ddf1}.  It
involves finding a set of weighting functions $w_i(\underline{u})$
such that, given an angular distribution $\sum_i b_i
f_i(\underline{u})$, where $\underline{u}$ is the vector of angular
variables,
\beq
E(w_i(\underline{u})) = b_i, 
\eeq
where $E$ stands for the expectation value.
Such a set of weighting functions
always exists \cite{ddf1} if the angular distribution is of the form
mentioned above.  An estimate of $b_i$ is then 
\beq
\hat{b}_i = \frac{1}{n} \sum_{j=1}^{n} w_i(\underline{u}_j)~~.
\eeq
The estimate is unbiased, i.e. $E(\hat{b}_i) = b_i$ and its standard
error is equal to $\sigma_i/\sqrt{n}$, where
\begin{eqnarray}
\sigma_i^2 & = &  \int (w_i(\underline{u}) - b_i )^2 \, f(\underline{u})
d\underline{u} \nonumber \\
& = & \left( \int (w_i(\underline{u}))^2 \,
  f(\underline{u}) d\underline{u} \right) \; - \; \left( \int
  w_i(\underline{u}) \, f(\underline{u}) d\underline{u} \right)^2 ~~.
\end{eqnarray}
The information in a sample of size $n$, on $b_i$ can be measured by
the inverse of the variance of $\hat{b}_i$ and is equal to
$n/\sigma_i^2$.  The information {\em per observation} is then
$1/\sigma_i^2$.  To compare the angular moments method with the
likelihood method we will compare the information per observation with
that of the likelihood method, as defined in the previous section.

Let us take the example of the transversity angle distribution without
any time information.  The density is given by (\ref{eq:only-angle}).
We shall see that in this case, the method of angular moments performs
almost as well as the maximum likelihood estimate in the region of
interest to us.

\begin{figure} [t]
\centerline{\psfig{figure=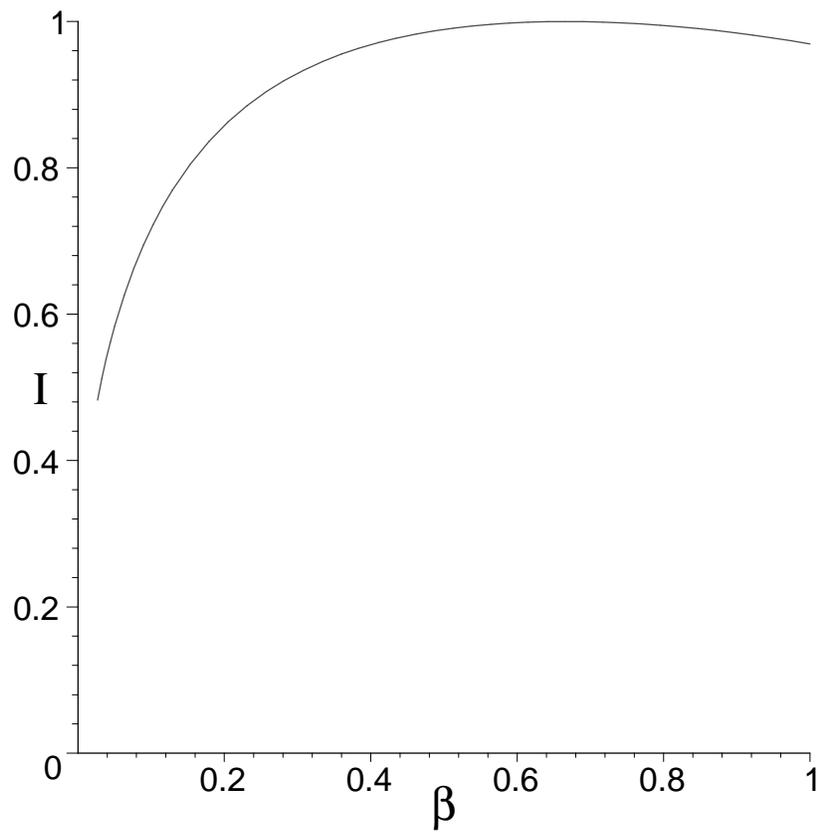,height=15cm}}
\caption{The ratio of information content about $\beta$ extracted
through the angular moments method and the maximal likelihood
method. The X-axis is the actual value of $\beta$.
\label{fig:ratio}}
\end{figure}

The information content in the maximal likelihood method is as given
in Eq.~(\ref{info-identity}). For the method of angular moments, the
density of $u=\cos(\theta)$ is 
$$p(u) = (3/8) [\beta(3u^2-1)+2(1-u^2)]~~.$$
The weighting function for $\beta$ may be chosen to be 
$w(u) = 5u^2 -1$, so that $E( w ) = \beta$ and $E(
w^2 ) = (24/7) \beta + (8/7) (1 - \beta)$.  The ratio
$$
I(\beta|u)_{AM} / I(\beta|u)_{ML}  
$$
(where AM represents the method of angular moments and ML represents 
the maximal likelihood method) is plotted in Figure~\ref{fig:ratio}.

The plot shows that for $\beta > 0.3$, 
the ratio of variances
is more than 0.9. The expected value of $\beta$ 
($ 0.8 < \beta < 1.0$) is well within this 
range. 
Thus, in the physical region
of interest, the method of angular moments seems to perform almost
as well as the maximal likelihood fit.

When we move to a higher number of parameters, the maximum likelihood
method will try to maximize the multidimensional likelihood function
and the complexity of the method increases rapidly with the number of
dimensions.  The dimension of the parameter space does not affect the
implementation of angular moment method.  Therefore it is useful, at
the least, as a method for providing good initial estimates.
Additionally, if it is as efficient compared to the full likelihood
method as the one-angle case suggests, it could render the maximum
likelihood method unnecessary.

\subsection{The effect of measurement discretization}

So long we have assumed that the data are measured to the maximum
precision available.  In practice, that is not the case and the data
are usually reported as the midpoint of the interval in which the
measurement actually fell.  For example, if we are measuring a random
variable $X$ to a precision $h$, that means that all observations
falling in the interval $( x^*-h/2, x^*+h/2]$ are reported as $x^*$.
The resulting discretization of measurements can lead to a systematic
bias in measurements, because instead of recording the random
variable $X$, we are recording the {\em derived} random variable
$X^*$ with the probability distribution given by
$${\rm P}(X^* = x^*) = \int_{x^*-h/2}^{x^*+h/2} f(x) dx.$$
Let us
compare the difference the means of the true and derived random
variable, i.e., $E(X)$ and $E(X^*)$.  It suffices to compare terms of
the form
$$
\int_{x^*-h/2}^{x^*+h/2} x f(x) dx \;\; \hbox{ and } \;\;
x^*\int_{x^*-h/2}^{x^*+h/2} f(x) dx. $$
Now,
\begin{eqnarray*}
\lefteqn{\int_{x^*-h/2}^{x^*+h/2} t f(t) dt - x^*\int_{x^*-h/2}^{x^*+h/2}
  f(t) dt}\\
  &= & \int_{x^*-h/2}^{x^*+h/2} ( t - x^*) f(t) dt \\
 & = &
  \int_{x^*-h/2}^{x^*+h/2} ( t - x^*) ( f(x^*) + (t-x^*) f^\prime(x^*)
  + O((t-x^*)^2) dt\\
 & \simeq & \int_{-h/2}^{+h/2} (t  f( x^*) + t^2 f^\prime(x^*)) dt =
  f^\prime(x^*) \frac{h^3}{6}.\\
\end{eqnarray*}
Thus, the error in discretization is of the order of the cube of the
length of the interval length of discretization if the density is
``well-behaved''.  If the (effective) support of the distribution of $X$ is
$[a,b]$, and the precision is $h$, then using a crude bound we would get
\begin{equation}
|E(X) - E(X^*)| \leq \frac{(b-a)}{h} \;\; \frac{h^3}{6} \left( \sup_{x
\in [a,b]} |f^\prime(x)| \right) = (b-a) \frac{h^2}{6} \left( 
\sup_{x \in [a,b]} |f^\prime(x)| \right)
\end{equation}
The bound has obvious modifications when the discretization is done
over intervals of varying length.

Thus the bias due to discretization is of the order of the square of
the bin widths.  If the bin widths are sufficiently small, neither the
moment method nor the likelihood method will be affected significantly.

\subsection{The effect of random error in measurements}
Another source of error in measurements comes from errors in the
measuring instruments.  Suppose the true variable we want to measure
is $X$, but instead, due to random error we measure 
$$Y = X + E,$$
where the error distribution has density $f_E(\cdot)$
and is independent of $X$.  It is reasonable to assume that the random
error has mean 0.  Suppose its variance is $\sigma^2_E$.  Then, $E(Y)
= E(X)$, but $V(Y) = V(X) + \sigma^2_E$.  In other words, the mean of
our measurements is unchanged by the random error, but the variance is
increased.  The implication is that the method of angular moments is
unaffected by random error as far as the estimation goes.  However,
the standard errors of the estimates will be increased.  

The effect of random error on the likelihood method is more serious,
because it relies on the {\em exact} mathematical form of the density of the
observed measurements.  For example, when we are measuring time and
the transversity angle, the density of the data will no longer be
(\ref{time-angle}) but
$$p(u,t|\beta, \Gamma_H, \Gamma_L) =
\int \int 
\left( \frac{3}{8} \beta \Gamma_L (1+u_*^2) e^{-\Gamma_L t_*} +
        \frac{3}{4}  (1 - \beta) \Gamma_H (1-u_*^2) e^{-\Gamma_H t_* }
        \right) \times $$
\begin{equation}
{\hphantom  {p(u,t|\beta, \Gamma_H, \Gamma_L) =
\int \int}} \times f_E( u-u_*, t-t_*) du_* dt_* 
\label{time-angle-error}
\end{equation}
In general, the likelihood function in the presence of random noise
is in the form of an integral with respect to the noise terms which
may be
analytically intractable unless the distribution of the noise term is
known and is in a simple form. 

If the noise term is believed to be significant, then it may be better
to use the method of moments because it is robust to the presence of
additional random noise.

\section{Three-angle distribution}
\label{three-ang}

        Given the enormous additional amount of information available
in the angular data $\theta$ as compared to the time data alone, 
we expect that the information embedded in the 
measurements of the two additional physical angles $\psi$ and $\varphi$     
would be useful in reducing the uncertainty on the parameters
which can in principle be measured by the time and transversity
angle data. Moreover, the additional terms available for
measurement in the three angle case [See Table~\ref{tab1}] 
allow us the access to additional parameters. The quantities  
$A_{\|}/A_0, ~A_{\perp}/A_0$ (both magnitudes and phases),
$\Delta m$ and the CP-violating parameter $\delta \phi$ need
the measurement of these two extra angles. The CP asymmetry
\beq
 (e^{-\Gamma_H t} - e^{-\Gamma_L t}) \cos(\delta_i) 
\delta  \phi
\label{untagged-asym}
\eeq
 can be measured even without tagging (without 
knowing whether the initial particle was a $B_s$ or 
$\overline{B_s}$) as long as we have this information. Using 
all the angular data, therefore, is highly recommended. Here,
we perform some  monte-carlo simulations to estimate 
how well the above parameters will be known in the next few years.

At the end of CDF run II (expected 
integrated luminosity $\approx 2~fb^{-1}$), 
we should have around 9000 fully
reconstructed  $B_s \to J/\psi (\to l^+ l^-) \phi (\to K^+ K^-)$
events \cite{cdf}, whereas this number is expected to 
increase by a factor
of at least 15 (just due to the integrated luminosity improvement) 
with TeV33. 
Sets of 10,000 and 100,000 events were generated, with 
 the accuracy 
in the measurements of time and angles taken at 
$\Delta t = 0.1/ \Gamma_L$ and $\Delta \theta = 0.005$.
The method of angular moments 
and time moments 
\footnote{ The $n^{th}$ time moment of a quantity $Q(t)$ is defined as
$T^{(n)} \equiv \int_0^{\infty} dt ~t^n ~Q(t)~~.$ Zeroth time
moment is just the time integrated quantity.}   
was used 
to recalculate all the
input parameters (with only the data, without any external information)
and histograms were plotted for the recalculated parameters. 
The simulations use the following set of parameters~: 
$$\frac{\Gamma_H}{\Gamma_L} = 0.8, 
~~|\frac{A_{\|}}{A_0}| = 0.55,  
~~|\frac{A_{\perp}}{A_0}| = 0.45, 
~~\frac{\delta m}{\overline{\Gamma}} = 10.0,
~\delta_1 = 0.5, ~\delta_2 = 2.5.$$ 

This choice of parameters is consistent with the corresponding ones
reported in \cite{cleo} for $B \to J/\psi K^{\ast}$ and
flavor $SU(3)$ (except for the lifetime difference).
It is seen that varying these parameters does not change the 
essential conclusions.

\begin{figure} [t]
\centerline{\psfig{figure=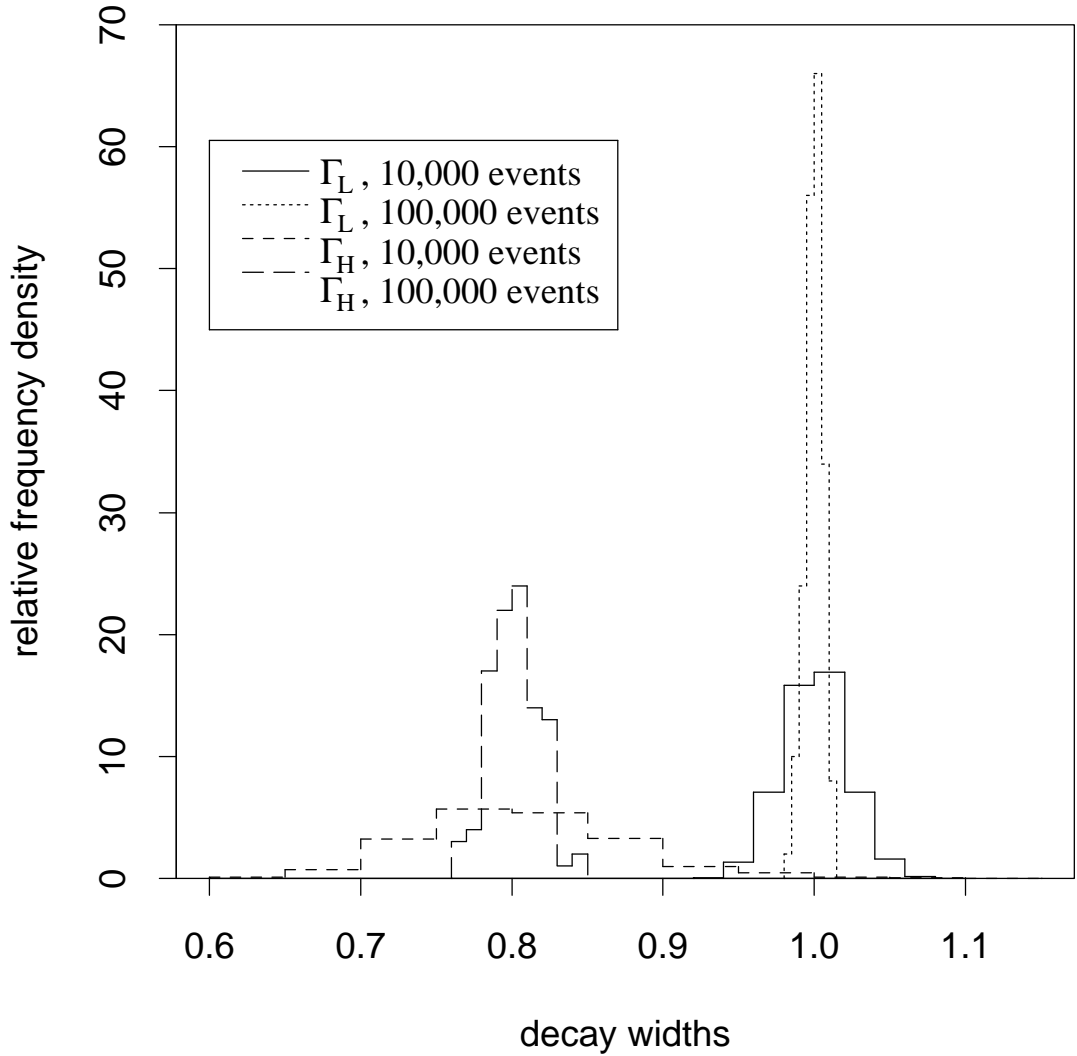,height=15cm}}
\caption{Determination of $\Gamma_H$ and $\Gamma_L$. The X-axis
has been normalized to $\Gamma_L \mbox{\it (actual)} = 1.0$.
\label{fig:gammas}}
\end{figure}

\begin{figure} [t]
\centerline{\psfig{figure=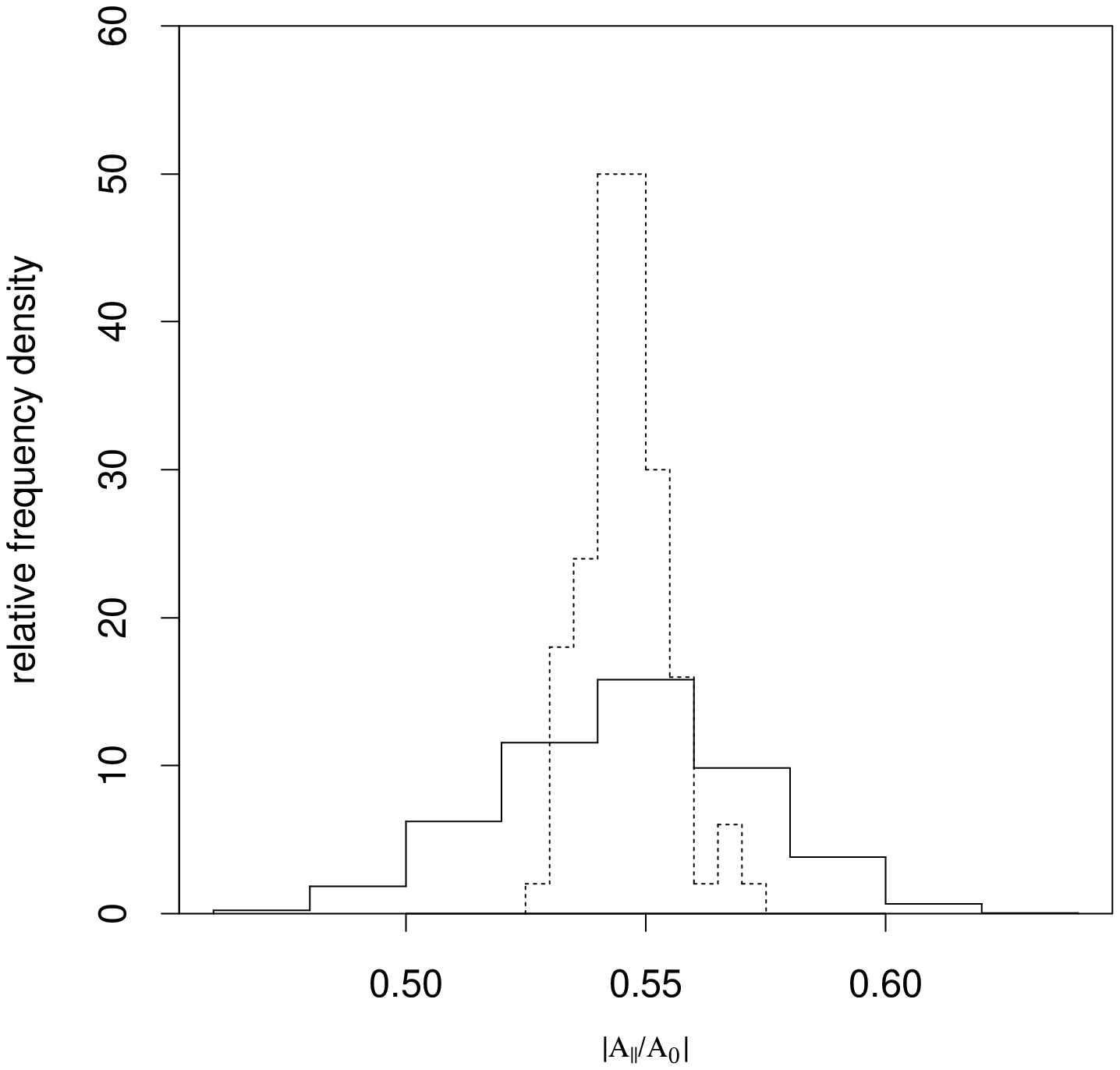,height=15cm}}
\caption{Determination of $|A_{\|}/ A_0|$. The solid line is for
        10,000 events and the dashed line is for 100,000 events.
\label{fig:apar}}
\end{figure}

\begin{figure} [t]
\centerline{\psfig{figure=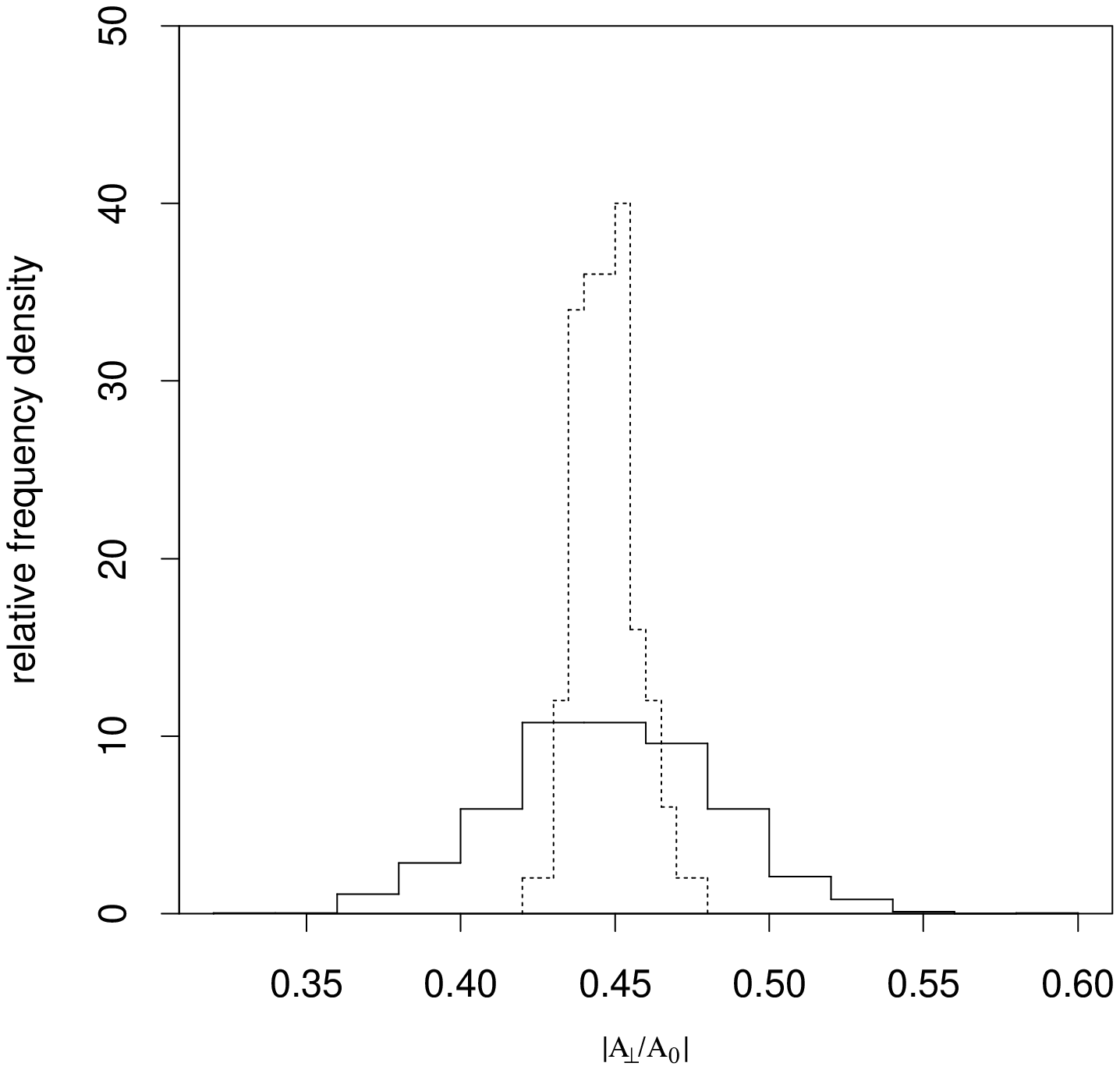,height=15cm}}
\caption{Determination of $|A_{\perp}/A_0|$.The solid line is for
        10,000 events and the dashed line is for 100,000 events.
\label{fig:aperp}}
\end{figure}

\begin{figure} [t]
\centerline{\psfig{figure=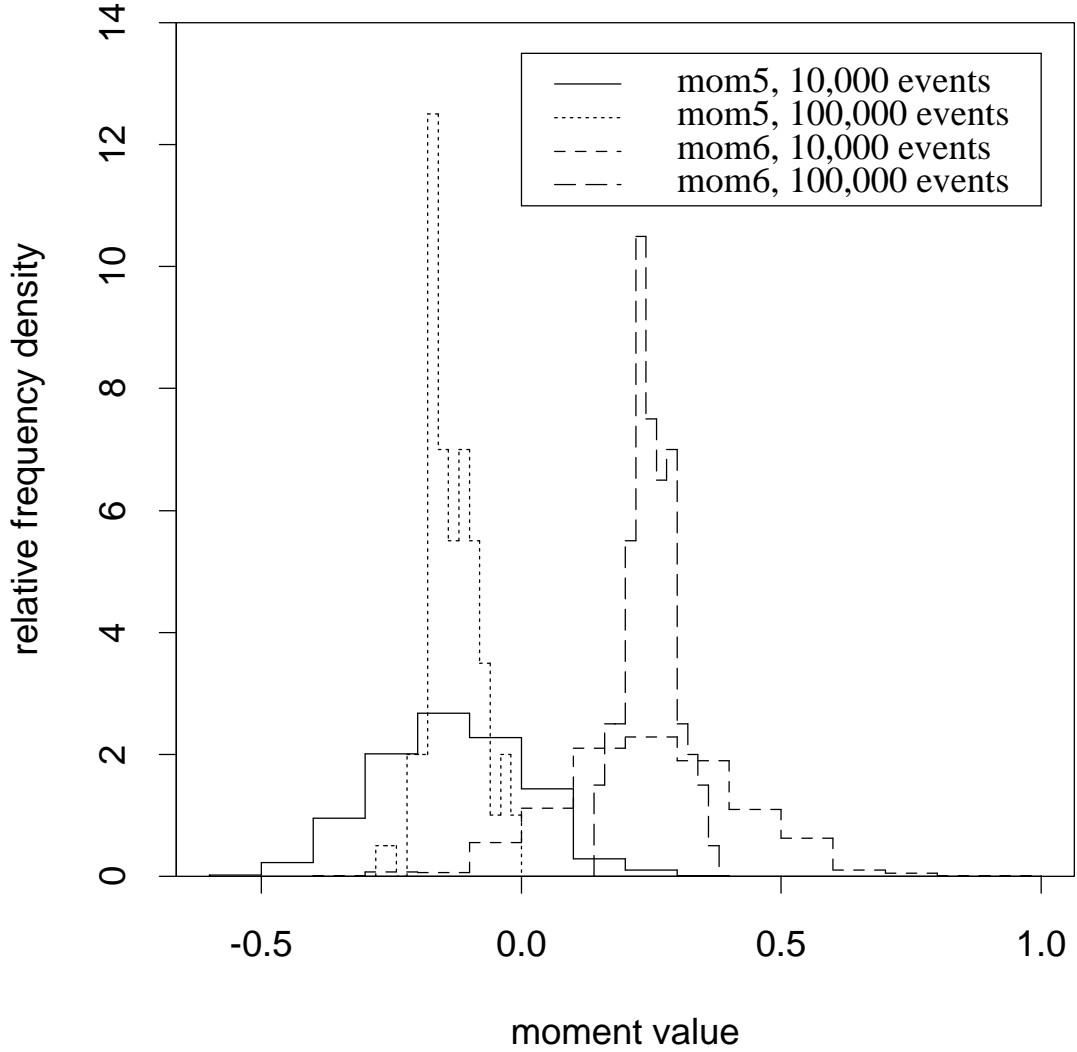,height=15cm}}
\caption{The moments of the ``Im'' terms in eq.~\ref{mom5}.
                mom5 is the value of
$-|A_{\|}||A_{\perp}|\cos(\delta_1+\kappa)\overline{\Gamma}/
\sin\kappa$
        and mom6 is the value of
$-|A_0||A_{\perp}|\cos(\delta_2+\kappa)\overline{\Gamma}/
\sin\kappa$.
\label{fig:moments}}
\end{figure}

Figures~\ref{fig:gammas}-\ref{fig:moments} show the results of 
these simulations. The Y-axis has been normalized to get
the `relative frequency density', such that the area under each
histogram is equal. The following observations should be noted.

\begin{itemize}
  
\item As can be seen from Fig.~\ref{fig:gammas}, the values of
  $\Gamma_H$ and $\Gamma_L$ are well-separated in the first stage
  (10,000 events) itself. With 100,000 events, the difference
  $\Gamma_L - \Gamma_H$ can be determined to nearly $\pm 0.05
  \Gamma_L$ to more than 95\% confidence level.  By virtue of the
  Central Limit Theorem, the method of moment estimates are
  approximately Gaussian in large samples.  The visual appearance of
  the histograms is consistent with this theoretical property.
  The width of the Gaussian distribution $\Gamma_L - \Gamma_H$ is then
  (approximately) inversely proportional to $\beta (1 - \beta)$ [See
  Sec.~\ref{sec:angular}] and has a weak dependence on the actual
  value of $\Gamma_L - \Gamma_H$ as long as $\Gamma_L - \Gamma_H$ is
  small, which is the case here.  So the above quantitative inferences
  from this histogram should stay valid even with a smaller value of
  $\Gamma_L - \Gamma_H$. Determination of $1 - \Gamma_H/\Gamma_L$ to
  0.05 is thus within reach. Even the small lifetime difference
  predicted recently in \cite{beneke} may be probed with this.

\item The accuracy in the measurements of $|A_{\|}/A_0|$, 
$|A_{\perp}/A_0|$ is as indicated in figures~\ref{fig:apar} and 
\ref{fig:aperp} respectively. The predictions of form factor 
models \cite{ffmodels} can thus be directly tested here.

\item The signs of $\cos(\delta_1)$ and $\cos(\delta_2)$ 
are important in order to resolve a discrete ambiguity in the 
CKM angle $\beta$, as pointed out recently \cite{ddf2}. 
In fact, if $\delta \phi$ is small
$(\approx 0.03)$ as predicted by the standard model, these
signs may be obtained without any time measurements
as follows.
With $\delta \phi$ neglected, the time integrated angular
moments of the ``Im'' terms in Table~\ref{tab1} give 
\beq
 - |A_X| |A_{\perp}| ~\cos(\delta_i + \kappa) ~\overline{\Gamma} 
/ \sin \kappa  ~~,
\label{mom5}
\eeq
 where 
$\kappa = \tan^{-1}(\overline{\Gamma}/\Delta m)$ and
$ X \in \{ 0, \| \}$. Since 
$\sin \kappa$ is positive, the sign of these moments immediately
give the sign of $\cos(\delta_i + \kappa)$, and given an upper
limit (of $\approx 0.1$) on the value of $\kappa$, will give
the sign of $\cos(\delta_i)$ as long as the value of this 
moment (\ref{mom5}) is not close to zero. Thus, just the 
sign of the angular moments of the ``Im'' terms in Table~\ref{tab1}
would be sufficient to resolve a discrete ambiguity in $\beta$. 
The relevant angular moments (time integrated) are shown in
Fig.~\ref{fig:moments}. 
The widths of these moment histograms
depend only weakly on the
actual parameter values and the plot can be used as a guide to
estimate the errors on the values of these moments for any other
parameter values.

\item  When $\Gamma_H \approx \Gamma_L$,
\beq
\int_0^{\infty} (e^{-\Gamma_H t} - e^{-\Gamma_L t}) \approx
        (\Gamma_L - \Gamma_H) / \overline{\Gamma}^2 ~~.
\eeq
The ability to measure  $\Gamma_L-\Gamma_H$, combined with 
the measurement of the time-integrated
CP asymmetry in Eq.~(\ref{untagged-asym}) (even without tagging)
would give a lower bound
on $\delta \phi$.
A high value of $\delta \phi$ would be a clear signal of
physics beyond the standard model.
In the next generation of experiments (TeV33 or LHC),
accurate values of $\delta_i ~~(i=1,2)$ will be obtained and 
$\delta \phi$ can be pinpointed.

\end{itemize}

 Feasibility studies for the measurement of $\Delta m / \Gamma$
and the asymmetries in 
this decay mode using the angular moments 
method and weighting functions have been made in \cite{galumian}
for the CMS detector, which claim that with $L \approx 10 fb^{-1}$,
 reasonable sensitivity on the oscillations
will be obtained at $\Delta m / \Gamma < 40$. 
The angular moments method has also been used for the analysis 
of $B^0 \to D^{*-} \rho^+$ and $B^+ \to \overline{D}^{*0} \rho^+$
\cite{dstar-rho} and the error estimation (Tables III and IV)
indicates that the angular moments method is almost as efficient
as the best fit method in estimating the observables, even with
the three angle distribution.

\section{Summary and Conclusions}
\label{concl}

Using a `reasonable' method for quantifying the information in the
data, we have shown that the information content in the data may
increase by orders of magnitude in the region of interest if angular
information is added to the time information. This is true even if the
quantity to be measured, e.g. the lifetime difference between $B_s^L$
and $B_s^H$, has no direct angular dependence. We have also isolated
`averaged' quantities for which this increase of information is small,
which means that their measurements are not helped much by the angular
data.

The actual use of the angular data involves the choice of a statistical
method to summarize the data.  The standard maximum likelihood method
is theoretically the ``best'' when the number of data points is very
large.  However, when the number of parameters to be estimated is
large, the numerical maximization of the likelihood may be difficult,
and if proper care is not taken, misleading results may be
obtained. 
Also, 
if there are
random errors in the measurement process, then 
the likelihood function would be
an integral that may not be mathematically known or, if known,
not evaluable in a closed form.

The method of angular moments is very straightforward to implement, and
the connections to the parameters to be determined are more transparent.  
It
is consistent in the statistical sense that, with infinite data, it
will nail the parameters down.  
Unlike the maximum
likelihood method, it is robust under random errors of measurement.  In
the one angle case at least, as we have shown explicitly, 
it is almost as efficient as the maximum likelihood
method in the region of interest.  Both methods are subject to
discretization errors which will be small if the interval of
discretization is small.  We therefore recommend the use of the
method of angular moments 
for extracting information, at least for the initial
estimates. If necessary, they can be refined with the likelihood method.
Even if the maximum likelihood method
is used, optimization routines require consistent starting values
which can be provided by the method of angular moments.

We have used the angular moments method on simulated sets of data to
estimate the accuracy to which it may determine the quantities of
interest.  In the case of the decay $B_s \to J/\psi (\to \ell^+
\ell^-) \phi (\to K^+ K^-)$, we find that in the first stage of
experiments (CDF II), this method should be sufficient to give
reliable values of $\Gamma_L - \Gamma_H$, $|A_{\|}/A_0|$ and
$|A_{\perp}/A_0|$. This, combined with the untagged CP asymmetry
measured through the same decay, would give a lower bound for $\delta
\phi$, which is expected to be very small in the standard model. The
signs of $\cos(\delta_1)$ and $\cos(\delta_2)$, which are useful in
resolving a discrete ambiguity in the CKM angle $\beta$, can be
determined in the next stage (TeV33), along with more accurate
determination of $\delta \phi$, which may point unambiguously to new
physics.

\section*{Acknowledgments}

We would like to thank the University of Chicago Physical
Sciences Division where this work was commenced, and was
supported in part by the United States Department of Energy
under Contract No. DE FG02 90ER40560.
A.D. would like to thank I. Dunietz, R. Fleischer, H. Lipkin, and
J. Rosner for previous collaborations on angular distributions
and for helpful comments on this manuscript,
S. Pappas for experimental insights, and K. Uryu for help in numerical
computations. S.S. would like to thank J. Liu for supporting this work
through National Science Foundation grants DMS-9101311 DMS-9501570 and
P. McCullagh and X.-L. Meng for helpful discussions.

\section*{Appendix}
\appendix

\section{Quantifying information in an experiment} 
\label{exp-info}
The notion of ``information'' that we have used in this paper is
derived from statistical theory.  A good reference is \cite{coxandh}.
Suppose an experiment is performed to determine the value of the parameter
$\alpha$.  The (average) information in the experiment to
discriminate between different possible values of the parameter when
the true value of the parameter is $\alpha_0$, is measured by
\beq
\label{info-identity}
I(\alpha_0,X) = -\int \ddot{\ell}(\alpha_0) \, p(x|\alpha_0) \;
dx~~, \eeq often called the {\em expected} Fisher information.  Here $X$ is
the random variable denoting the data used from the experiment,
$p(x|\alpha)$ is the probability
of $X$ given the parameter value $\alpha$, and $\ell(\alpha) = \log(
p(x|\alpha) )$ is the log likelihood function. Note the dependence of
the expected information in the experiment on the true value of the
parameter, $\alpha_0$ and on the data used, $X$.  Both $\alpha$ and
$X$ may be vector-valued.  This measure of information possesses the
additivity property, i.e. if $X_1$ and $X_2$ denote data from two
independent experiments about the same parameters, then
$${\rm I}(\alpha_0,(X_1,X_2)) = {\rm I}(\alpha_0,X_1) + {\rm
  I}(\alpha_0,X_2)~~.$$
In particular this implies that if $n$
independent and identically distributed data points, $X_1, X_2,
\ldots, X_n$, are collected from an experiment, the expected
information in the whole experiment is $n$ times the expected
information in one observation.
$$
{\rm I}(\alpha_0,(X_1,X_2,\ldots,X_n)) = n \, {\rm
  I}(\alpha_0,X_1).$$
Additionally, it can be shown that for any estimator of $\alpha$, say
$\hat{\alpha}_n$, based on a sample $X_1, X_2, \ldots, X_n$, (the
Cram\'er-Rao inequality)
\begin{equation}
\label{cramer-rao}
V(\hat{\alpha}_n) \geq \frac{1}{{\rm
    I}(\alpha_0,(X_1,X_2,\ldots,X_n))} = \frac{n}{{\rm
    I}(\alpha_0,X_1)} ~~,
\end{equation}
where $V$ is the variance and $\hat{\alpha}_n$ is based on a sample of
size $n$.
When the sample size is large and certain regularity conditions hold,
the lower bound in the variance is achieved by the maximum likelihood
estimate.  It is in this sense that the maximum likelihood estimate is
the ``best''.  

It can also be shown that when we have independent and identically
distributed data points, for large samples,
\beq
\ell(\alpha) \simeq \ell(\widehat{\alpha}) + (\alpha - \widehat{\alpha})
\dot{\ell}(\widehat{\alpha}) + (\alpha - \widehat{\alpha})^2
\ddot{\ell}(\widehat{\alpha}) 
\eeq
where $\widehat{\alpha}$ denotes the maximum likelihood estimate of
$\alpha$.  By construction, $\dot{\ell}(\hat{\alpha})=0$ and hence 
\beq
 \ell(\alpha) \simeq \ell(\widehat{\alpha}) + (\alpha - \widehat{\alpha})^2
\ddot{\ell}(\widehat{\alpha}).
\eeq
In other words, the log likelihood surface is approximately quadratic and
its shape can be described by the position of the maximum
($\widehat{\alpha}$) and the curvature of the log likelihood in the
neighbourhood of the maximum ($\ddot{\ell}(\widehat{\alpha})$).  The
latter describes how fast the log likelihood falls off; the larger the
value of $\ddot{\ell}(\widehat{\alpha})$, the steeper the fall and
stronger is the evidence in favour of values near the maximum.  For
this reason, $\ddot{\ell}(\widehat{\alpha})$, is also used as a
measure of information, but since it varies from sample to sample, it
is called the {\em observed} Fisher information.  Its average value
is the {\em expected} Fisher information mentioned above.

While we have defined the information for a scalar parameter, the
general idea can be extended to vector-valued parameters.  When there
are two or more parameters, the appropriate measure of expected information is
the expected information matrix which is the expected value of the
hessian of the log likelihood as in (\ref{info-identity}).  See
\cite{coxandh} for details and additional references.

\newpage

\centerline{\large Table Captions}
\bigskip

{\bf Table 1. ~~:~~} 
Time evolution of the decay $B_s\to J/\psi(\to l^+l^-)
\phi(\to K^+K^-)$ of an initially (i.e.\ at $t=0$) pure $B_s$ meson.

\newpage

\centerline{\large Figure Captions}
\vglue 0.4cm
\noindent

{\bf Fig. 1.~~:~~} 
The ability to guess the decay type in four scenarios:
      Plots of the {\em posterior} probability that a decay is of the first
    type (with mean lifetime $1/\Gamma_L$) given 
    \begin{list}{$\bullet$}{ \setlength{\itemsep}{-\parsep} 
    \setlength{\topsep}{2pt} }
    \item only angular information, $u=\cos(\theta)$, (solid line)
    \item only time data, $\Gamma_H/\Gamma_L=1$, (dotted line)
    \item only time data, $\Gamma_H/\Gamma_L=1.2$, (narrowest dashed
      line)
    \item only time data, $\Gamma_H/\Gamma_L=20$. (broad dashed line)   
        \end{list}

\bigskip

{\bf Fig. 2.~~:~~}
The ratio of the variances,
$V(\hat{\theta}_i(t)) / V(\hat{\theta}_i(u,t))$ of the estimates of
$\theta_1 = \beta \Gamma_L + (1-\beta) \Gamma_H$.

\bigskip

{\bf Fig. 3.~~:~~}
The ratio of the variances,
$V(\hat{\theta}_i(t)) / V(\hat{\theta}_i(u,t))$ of the estimates of
$\theta_2 = \Gamma_L - \Gamma_H$.

\bigskip

{\bf Fig. 4.~~:~~}
The ratio of the variances,
$V(\hat{\theta}_i(t)) / V(\hat{\theta}_i(u,t))$ of the estimates of
$\theta_3 = \beta$.

\bigskip

{\bf Fig. 5.~~:~~}
The ratio of information content about $\beta$ extracted
through the angular moments method and the maximal likelihood
method. The X-axis is the actual value of $\beta$.

\bigskip

{\bf Fig. 6.~~:~~}
Determination of $\Gamma_H$ and $\Gamma_L$. The X-axis
has been normalized to $\Gamma_L \mbox{\it (actual)} = 1.0$.

\bigskip

{\bf Fig. 7.~~:~~}
Determination of $|A_{\|}/ A_0|$. The solid line is for
        10,000 events and the dashed line is for 100,000 events.

\bigskip

{\bf Fig. 8.~~:~~}
Determination of $|A_{\perp}/A_0|$.The solid line is for
        10,000 events and the dashed line is for 100,000 events.

\bigskip

{\bf Fig. 9.~~:~~}
The moments of the ``Im'' terms in eq.~\ref{mom5}.
                mom5 is the value of
$-|A_{\|}||A_{\perp}|\cos(\delta_1+\kappa)\overline{\Gamma}/
\sin\kappa$
        and mom6 is the value of
$-|A_0||A_{\perp}|\cos(\delta_2+\kappa)\overline{\Gamma}/
\sin\kappa$.

\newpage

\begin{table}[t]
\begin{center}
\begin{tabular}{|c|l|}
\hline
Observables & Time evolutions \\
\hline
$|A_0(t)|^2$  & $|A_0(0)|^2 \left[e^{-\Gamma_L t} -
e^{-\overline{\Gamma}t}
\sin(\Delta m t)\delta\phi\right]$\\
$|A_{\|}(t)|^2$ &$ |A_{\|}(0)|^2 \left[e^{-\Gamma_L t} -
e^{-\overline{\Gamma}t}\sin(\Delta m t)\delta\phi\right]$\\
$|A_{\perp}(t)|^2$ & $|A_{\perp}(0)|^2 \left[e^{-\Gamma_H t} +
e^{-\overline{\Gamma}t}\sin(\Delta m t)\delta\phi\right]$\\
\hline
Re$(A_0^*(t) A_{\|}(t))$ &  $|A_0(0)||A_{\|}(0)|\cos(\delta_2 -
\delta_1)\left[e^{-\Gamma_L t} - e^{-\overline{\Gamma}t}
\sin(\Delta m t)\delta\phi\right]$\\
Im$(A_{\|}^*(t)A_{\perp}(t))$ & $|A_{\|}(0)||A_{\perp}(0)|\left[
e^{-\overline{\Gamma}t}\sin (\delta_1-\Delta m t)+\frac{1}{2}\left(
e^{-\Gamma_H t}-e^{-\Gamma_L
t}\right)\cos(\delta_1)\delta\phi\right]$\\
Im$(A_0^*(t)A_{\perp}(t))$ & $|A_0(0)||A_{\perp}(0)|\left[
e^{-\overline{\Gamma}t}\sin (\delta_2-\Delta m t)+\frac{1}{2}\left(
e^{-\Gamma_H t}-e^{-\Gamma_L
t}\right)\cos(\delta_2)\delta\phi\right]$\\
\hline
\end{tabular}
\end{center}\caption{Time evolution of the decay $B_s\to J/\psi(\to l^+l^-)
\phi(\to K^+K^-)$ of an initially (i.e.\ at $t=0$) pure $B_s$ meson.}
\label{tab1}
\end{table}

\newpage

\end{document}